\documentclass[12pt]{article}
\usepackage{graphicx}
\usepackage{multirow}
\usepackage{booktabs}
\RequirePackage{xspace}
\usepackage{relsize}
\def\cm{\ensuremath{{\rm \,cm}}\xspace}
\def\mm{\ensuremath{{\rm \,mm}}\xspace}
\def\nm{\ensuremath{{\rm \,nm}}\xspace}
\def\sec{\ensuremath{\rm {\,s}}\xspace}

\def\cbar{\ensuremath{\overline c}\xspace}
\def\Dbar{\kern 0.2em\overline{\kern -0.2em D}{}\xspace}

\def\CP{\ensuremath{C\!P}\xspace}
\mathchardef\Upsilon="7107

\def\order{{\ensuremath{\cal O}}\xspace}

\def\invfb{\ensuremath{\mbox{\,fb}^{-1}}\xspace}
\def\invab{\ensuremath{\mbox{\,ab}^{-1}}\xspace}
\def\KS{\ensuremath{K^0_{\scriptscriptstyle S}}\xspace} 
\def\KL{\ensuremath{K^0_{\scriptscriptstyle L}}\xspace} 
\def\Ds{\ensuremath{D^+_s}\xspace}
\def\babar{\mbox{\slshape B\kern-0.1em{\smaller A}\kern-0.1em
    B\kern-0.1em{\smaller A\kern-0.2em R}}}
\newcommand{\etapr}{\ensuremath{\eta^{\prime}}\xspace}

\def\pbnr{}
\def\speaker{Gagan B. Mohanty}
\def\onbehalfof{the Belle II Collaboration}
\def\title{Charm at Belle II -- Status and Prospects}
\def\affiliation{Tata Institute of Fundamental Research, Homi Bhabha Road,\\
Colaba, Mumbai 400 005, India}
\def\support{Supported in part by the Department of Atomic Energy, India and a Grant-in-Aid for innovative scientific research area {\em Elucidation of new hadrons with a variety of flavors} from KEK, Japan.}

\newcommand\pubnumber{\pbnr}
\newcommand\pubdate{\today}
\def\Title#1{\begin{center} {\Large #1 } \end{center}}
\def\Author#1{\begin{center}{ \sc #1} \end{center}}

\newcommand{\OnBehalf}[1]{\sbox0{#1}\ifdim\wd0=0pt
        {}
	\else
	{\\on behalf of #1}
	\fi}
\newcommand{\SupportedBy}[1]{\sbox0{#1}\ifdim\wd0=0pt
        {}
	\else
	{\footnote{#1}}
	\fi}
\def\Address#1{\begin{center}{ \it #1} \end{center}}

\newcommand\pubblock{\includegraphics[width=5cm]{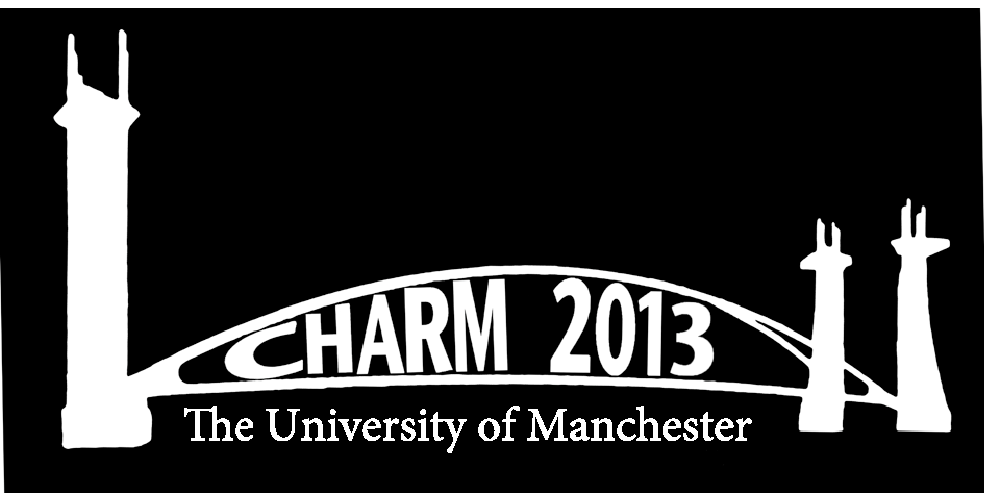}\hfill{\begin{tabular}{l} \pubnumber\\
         \pubdate  \end{tabular}}}
\newenvironment{Abstract}{\begin{quotation}  }{\end{quotation}}
\newenvironment{Presented}{\begin{quotation} \begin{center} 
             PRESENTED AT\end{center}\bigskip 
      \begin{center}\begin{large}}{\end{large}\end{center} \end{quotation}}
\def\Acknowledgements{\bigskip  \bigskip \begin{center} \begin{large}
             \bf ACKNOWLEDGEMENTS \end{large}\end{center}}
\def\venue{The 6$^{th}$ International Workshop on Charm Physics\\
(CHARM 2013)\\
Manchester, UK,  31 August -- 4 September, 2013}

\def\beq{\begin{equation}}
\def\eeq#1{\label{#1}\end{equation}}
\def\eeqn{\end{equation}}

\def\beqa{\begin{eqnarray}}
\def\eeqa#1{\label{#1}\end{eqnarray}}
\def\eeqan{\end{eqnarray}}

\let\bar=\overbar

\def\Dslash{\not{\hbox{\kern-4pt $D$}}}
\def\dslash{\not{\hbox{\kern-2pt $\del$}}}

\def\msb{{\bar{\ssstyle M \kern -1pt S}}}

\begin{document}
\begin{titlepage}
\pubblock

\vfill
\Title{\title}
\vfill
\Author{\speaker\SupportedBy{\support}\OnBehalf{\onbehalfof}}
\Address{\affiliation}
\vfill
\begin{Abstract}
High-precision flavor physics measurements play a complementary role to
the direct searches for new physics by CMS and ATLAS experiments at LHC.
Such measurements will be performed with the Belle II detector at the
upgraded KEKB accelerator (SuperKEKB) in Japan. The physics potential
with emphasis on the charm sector, current status and future prospects
of the Belle II experiment are presented in these proceedings.
\end{Abstract}
\vfill
\begin{Presented}
\venue
\end{Presented}
\vfill
\end{titlepage}
\def\thefootnote{\fnsymbol{footnote}}
\setcounter{footnote}{0}
%

\section{Introduction}

With the discovery of a Higgs-like boson~\cite{Higgs} by CMS~\cite{cms}
and ATLAS~\cite{atlas} experiments at the LHC, the standard model (SM) of
elementary particles is on the verge of completion. However, we still have
many compelling reasons to believe that it cannot be the whole story.
Notable among them are a ten orders of magnitude difference between the
matter-antimatter asymmetry in universe and the $\CP$ violation content of
the SM~\cite{hou}, and the absence of a suitable candidate to explain dark
matter~\cite{rubin}. Therefore, it is imperative to search for new physics
(NP) using a diverse and complementary set of probes that include
experiments at the energy frontier (CMS and ATLAS), the luminosity frontier
(LHCb~\cite{lhcb} and Belle II~\cite{belle2}), and the cosmic frontier
studying neutrinos, gamma-ray photons and cosmic rays.

Now one could ask, with the LHC running at full swing, whether there is
any need for an experiment operating on a low-energy $e^+e^-$ machine
(viz., Belle II at SuperKEKB). Moreover, if the goal is to do flavor
physics can we not just focus on LHCb alone? Well, we have good reasons
on both the counts. First, a flavor factory mostly studies processes
that occur at one-loop level in the SM but may be of $\order(1)$ in
various NP models. These processes include flavor-changing neutral current
(FCNC) decays, neutral meson mixing and $\CP$ violation in the decays
of beauty and charm hadrons. Being mostly associated with quantum loops,
they probe energy scales that cannot be directly accessed at the LHC.
Further, in case the LHC finds a NP signal, e.g., supersymmetry, the
flavor factory can play an essential role on deciphering its nature
through a systematic study of various flavor-violating couplings.
Back to the question LHCb {\it vs.} Belle II, thanks to its pristine
$e^+e^-$ environment, high trigger efficiency, and excellent photon
and $\pi^0$ reconstruction capabilities, the latter will have an edge
over LHCb in the study of final states comprising neutral particles and
missing neutrinos. Indeed, both Belle II (once operational) and LHCb
will work in tandem to explore the NP landscape providing a synergy to
the energy-frontier experiments.

In these proceedings, we discuss about the Belle II experiment to be
located at the SuperKEKB collider of Japan, which aims at collecting
$50$ times as much the data as Belle~\cite{belle} did few years ago. A
particular emphasis is placed on its scope of studies of charm hadrons
as a NP probe.

\section{Charm as a NP probe}

Charm decays via loop processes provide an interesting test bed for NP as
SM footprints for these decays are tiny due to GIM suppression~\cite{gim}
and the lack of a large hierarchy in the down-type quark masses. Possible
NP contributions can make their presence felt in FCNC processes that are
larger for the up-type than the down-type quarks. Charm decays are thus
best suited to reveal potential non-SM dynamics.

An important avenue that will be explored by Belle II in its pursuit of
NP is the search for $\CP$ violation in charm hadron decays. The most
obvious candidates here are the singly Cabibbo-suppressed decays~\cite{scs},
where the typical SM value for the $\CP$ asymmetry ($A_{\CP}$) is in the
range of $10^{-3}$. Of course, while talking about such a small effect
one would need a good control over the SM predictions, something that
is in general lacking in this sector because of long-distance effects.

In Table~\ref{tab:dcpv} we summarize results on the direct $\CP$
asymmetry for a number of charm decay modes from Belle. As we have tried
to categorize them in two horizontal blocks, LHCb has an upper hand for
the final states containing charged tracks, or neutral mesons that
can be reconstructed in charged-track final states {\it e.g.}, $\KS\to
\pi^+\pi^-$. On the other hand, Belle II will be very competitive for
the final states containing $\pi^0$, $\eta$ or $\etapr$. In fact, for
some of the channels such as $D^0\to\pi^0\pi^0$ it will be the only one
able to perform such measurement, thanks to its clean $e^+e^-$ environment.

\begin{table}[!ht]
\begin{center}
\caption{Summary of $\CP$ asymmetry measured by Belle for
various charm decay channels together with the size of the data
sample used (see the second and third columns) and the corresponding
projection for Belle II.}
\label{tab:dcpv}
\begin{tabular}{lccc}
\hline\hline
Channel & ${\cal L}_{\rm int}[\invfb]$ & $A_{\CP}[\%]$ & Belle II with $50\invab[\%]$\\
\hline
$D^0\to\KS\pi^0$ & $791$ & $-0.28\pm0.19\pm0.10$ & $\pm0.05$\\
$D^0\to\KS\eta$ & $791$ & $+0.54\pm0.51\pm0.16$ & $\pm0.10$\\
$D^0\to\KS\etapr$ & $791$ & $+0.98\pm0.67\pm0.14$ & $\pm0.10$\\
$D^0\to\pi^0\pi^0$ & $976$ & $\sigma[A_{\CP}]\approx 0.6\%$ & \\
$D^0\to\pi^+\pi^-\pi^0$ & $532$ & $+0.43\pm1.30$ & \\
$D^0\to K^+\pi^-\pi^0$ & $281$ & $-0.6\pm5.3$ & \\
$D^+\to\eta\pi^+$ & $791$ & $+1.74\pm1.13\pm0.19$ & $\pm0.20$\\
$D^+\to\etapr\pi^+$ & $791$ & $-0.12\pm1.12\pm0.17$ & $\pm0.20$\\
\hline
$D^0\to\pi^+\pi^-$ & $976$ & $+0.55\pm0.36\pm0.09$ & $\pm0.07$\\
$D^0\to K^+K^-$ & $976$ & $-0.32\pm0.21\pm0.09$ & $\pm0.05$\\
$D^0\to K^+\pi^-\pi^+\pi^-$ & $281$ & $-1.8\pm4.4$ & \\
$D^+\to\phi\pi^+$ & $955$ & $+0.51\pm0.28\pm0.05$ & $\pm0.05$\\
$D^+\to\KS\pi^+$ & $977$ & $-0.363\pm0.094\pm0.067$ & $\pm0.05$\\
$D^+\to\KS K^+$ & $977$ & $+0.08\pm0.28\pm0.14$ & $\pm0.10$\\
$\Ds\to\KS\pi^+$ & $673$ & $+5.45\pm2.50\pm0.33$ & $\pm0.30$\\
$\Ds\to\KS K^+$ & $673$ & $+0.12\pm0.36\pm0.22$ & $\pm0.10$\\
\hline\hline
\end{tabular}
\end{center}
\end{table}

Another interesting NP probe for Belle II is the rare FCNC and
forbidden decays of charm mesons. Under the first category will
be the channels $D^0\to\gamma\gamma$ and $D^0\to h\ell^+\ell^-$
($h=\pi^0,\eta,\omega$ and $\ell=e,\mu$). Going by the SM
predictions, even some of them could be observed for the first
time. Among the decays that are not allowed in the SM because of
lepton-flavor or lepton-number violation, Belle II will have a
better sensitivity for $D^0\to h\ell^{\pm}\ell^{\prime\mp}$
($h=\pi^0,\eta,\omega$ and $\ell,\ell^{\prime}=e,\mu$)
compared to LHCb. It can also probe decays with charged-track
final state such as $D^0\to e^{\pm}\mu^{\mp}$.

In addition to having almost two orders of magnitude larger data
compared to Belle and $\babar$, Belle II will see lots of innovation
in the data analysis techniques. One such nice idea is the so-called
{\em Charm tagging} where a recoiling charm hadron can be measured in
the continuum process $e^+e^-\to c\cbar\to\Dbar_{\rm tag}
X_{\rm frag}D^{(*)}_{\rm rec}$ just by reconstructing the tagged
$D$ meson, $\Dbar_{\rm tag}$, together with some fragmentation products
$X_{\rm frag}$. We have already some good working examples from the
current-generation $B$ factories; a recent one being the measurement
of absolute branching fractions of leptonic and hadronic $\Ds$ meson
decays at Belle~\cite{anze}. This idea will be explored further at
Belle II.

\section{KEKB to SuperKEKB}

To look for the possible deviation from SM predictions in the
flavor sector, at least two orders of magnitude larger data sample
in excess of $50\invab$ is required. Such a rise in the integrated
luminosity calls for an equally large increase in the instantaneous
luminosity,
\begin{eqnarray}
{\cal L}=\frac{\gamma_{e^{\pm}}}{2er_e}\left(1+\frac{\sigma^{\star}_y}{\sigma^{\star}_x}\right)\frac{I_{e^{\pm}}\xi^{e^{\pm}}_y}{\beta^{\star}_y}\,\frac{R_L}{R_{\xi_y}},
\label{lumi}
\end{eqnarray}
where $\gamma$ is the Lorentz factor, $e$ is the electron charge,
$r_e$ is the classical electron radius, $\sigma^{\star}_{x,y}$ and
$\beta^{\star}_y$ are the transverse beam size and vertical beta
function at the interaction point (IP), $I$ is the beam current,
$\xi_y$ is the beam-beam parameter, and $R_L/R_{\xi_y}$ is a
geometric factor because of finite beam-crossing angle and hour
glass effect. The $e^{\pm}$ refers to the product of the
corresponding quantities for low-energy positron (LER) and
high-energy electron (HER) beams.

KEKB holds the current world record for instantaneous luminosity which
is $2.1\times10^{34}\cm^{-2}\sec^{-1}$. In order to achieve a value
$40$ times larger that would enable us to accumulate the desired data
size, we need to improve on some of the parameters given
in Eq.~(\ref{lumi}). The main contribution comes from a significantly
smaller beam size at the IP. The $\beta^\star$ values are reduced from
$5.9\mm$ to $0.27/0.30\mm$ in the $y$ direction and from $1200\mm$ to
$32/25\mm$ in the $x$ direction for HER/LER. To keep the beam-beam
parameter that is proportional to $\sqrt{\beta^\star/\varepsilon}$ at
a similar level as KEKB, the emittance $\varepsilon$ is reduced from
$24/18\nm$ to $4.6/3.2\nm$ for HER/LER. This is accomplished by
installing a new electron source and a new damping ring, and by
redesigning the HER arcs. The last contribution to the luminosity
gain comes from the higher beam currents. They are increased from
$1.64/1.19$\,A to $3.6/2.6$\,A by modifying the radio-frequency
systems. Among other important changes of the KEKB complex are:
installation of TiN-coated beam pipe with antechambers, a completely
revamped interaction region having new superconducting/permanent final
focusing quads (QCS), and replacement of short dipoles with longer ones
for LER.

\section{Enter Belle II}

The steep increase in luminosity requires a careful design of various
detector components. The main issue will be how to mitigate the effect
of higher beam backgrounds (by a factor of $10$ to $20$), leading to
an increase in occupancy and radiation damage as well as fake hits and
pileup noise in the calorimeter. The higher event rate will also call
for a substantial modification in the trigger scheme, DAQ and computing
compared to Belle. Furthermore, to fully exploit the physics potential
of the experiment an improved hadron and muon (especially at low momentum)
identification capability, and a better hermiticity are required.

\begin{figure}[htb]
\centering
\includegraphics[width=0.9\columnwidth]{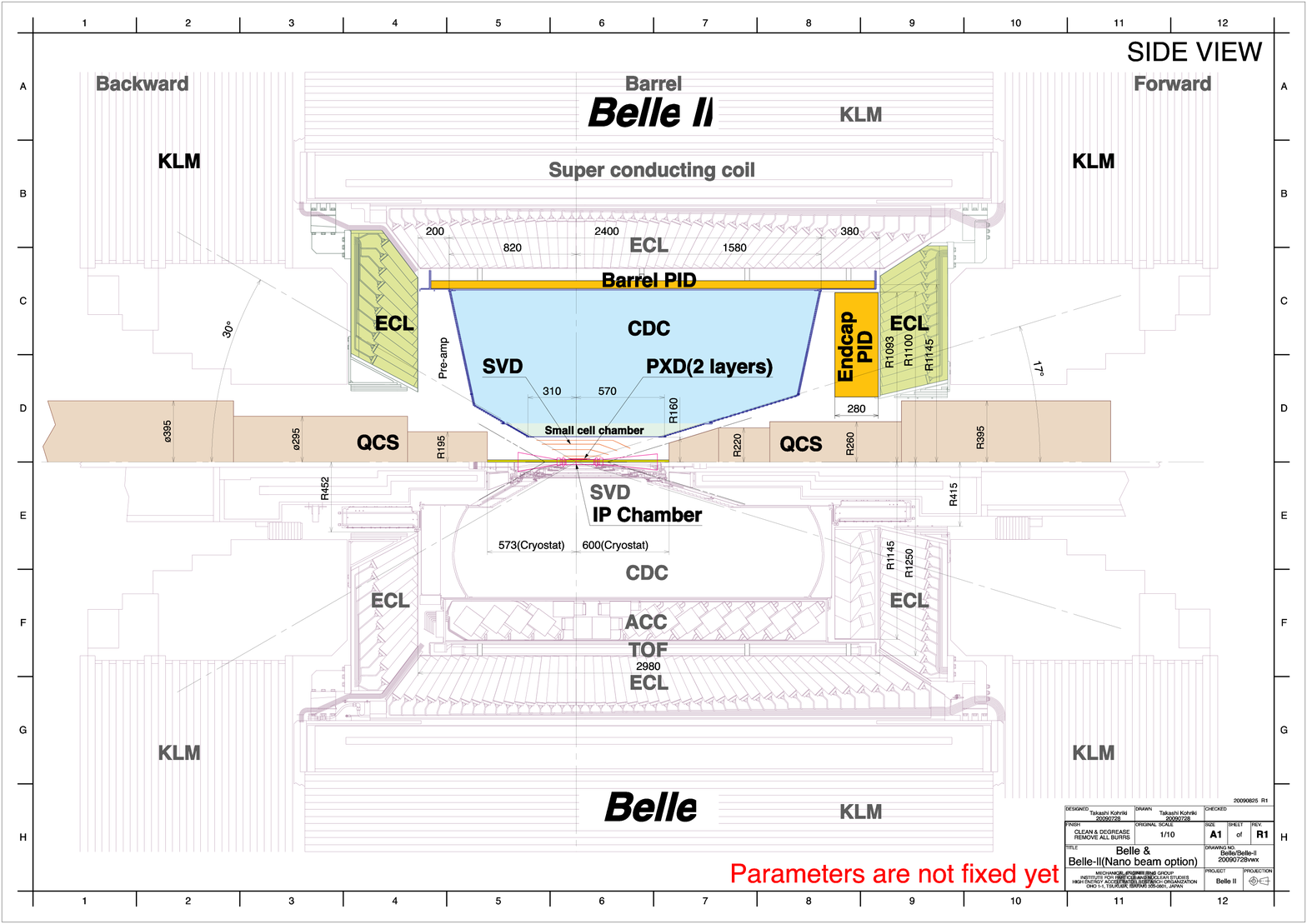}
\caption{Belle II (top half) compared to its predecessor, Belle
(bottom half).}
\label{belle2}
\end{figure}

Belle II -- almost a new experiment rather than being a refurbished
Belle -- will adopt the following solutions. The inner layers of the
vertex detector (VXD) will be replaced with a pixel detector, the main
tracking device (central drift chamber, CDC) will be augmented by an
inner silicon strip detector, a better charged hadron identification
(PID) device will be used, the CsI(Tl) crystals of the endcap calorimeter
(ECL) will be eventually replaced by pure CsI, the resistive plate chambers
of the endcap muon and $\KL$ detection system (KLM) will be replaced with
plastic scintillators read out by silicon photomultipliers, and all
detector components will be read-out by fast electronics and an improved
computing system. Figure~\ref{belle2} provides a comparison between Belle
and Belle II. Below we have picked out two specific systems (VXD
and PID), which will play a crucial role in probing NP, for further
deliberation.

\begin{figure}[htb]
\centering
\includegraphics[width=0.38\columnwidth]{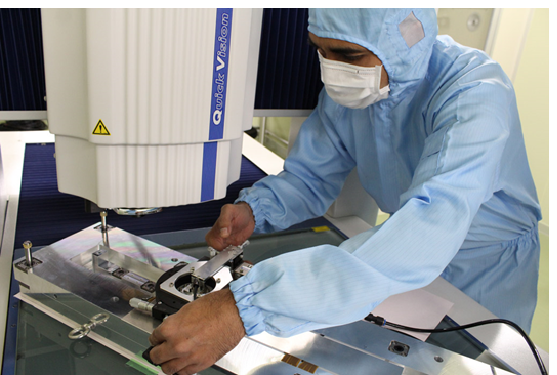}
\includegraphics[width=0.57\columnwidth]{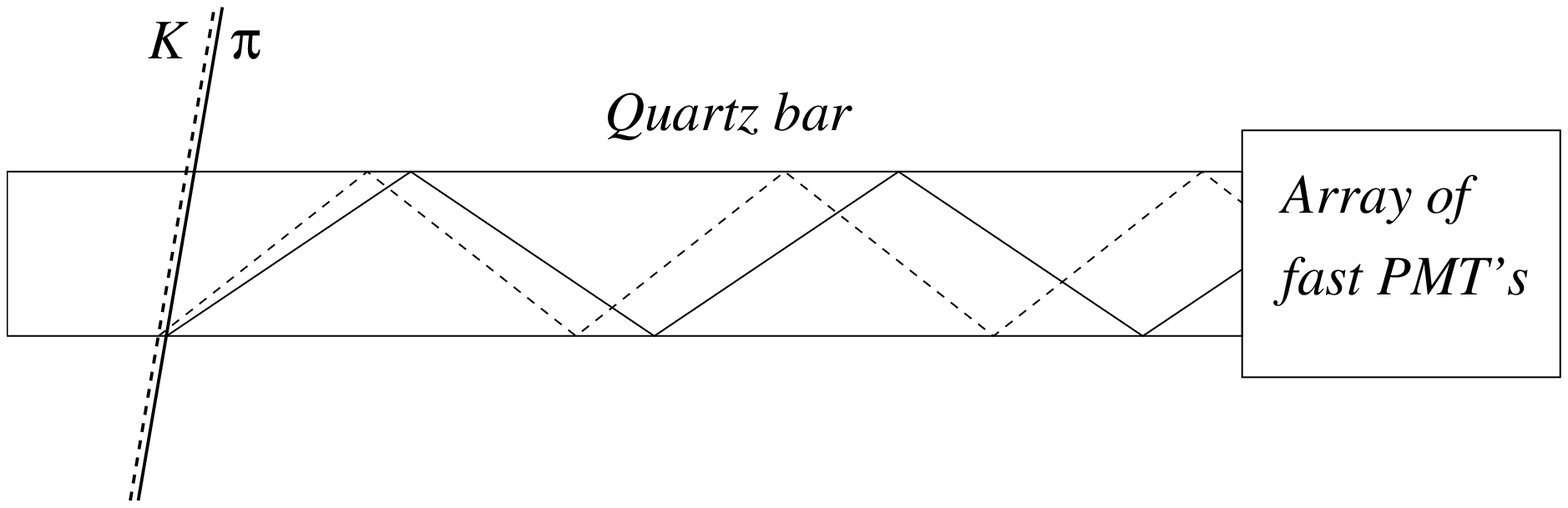}
\caption{(left) A part of the Origami assembly exercise for the SVD,
and (right) the principle of operation of the TOP counter.}
\label{detect2}
\end{figure}

The VXD will have two pixel layers (PXD) at $r=14\mm$ and $22\mm$ around
a $10\mm$-radius beryllium beam pipe, and four layers of double-sided
silicon strip sensors (SVD) at radii of $38\mm$, $80\mm$, $104\mm$, and
$135\mm$. The PXD detector will be based on DEPFET sensors whereas the
SVD will employ a novel chip-on-sensor readout scheme called
{\em Origami}~\cite{origami} in the outermost three layers. The latter
provides a low-mass solution for the double-sided readout (the left plot
in Fig.~\ref{detect2} shows a part of the Origami assembly exercise for
the SVD). A significant improvement in the vertex resolution, by a factor
of two, is expected both for low momentum particles owing to reduced
multiple scattering, and for high momentum particles because the
high-resolution pixel detector is closer to the IP. Another salient
feature is a better $\KS$ reconstruction efficiency (about $30\%$ more)
because of a larger volume covered by the VXD system.

The PID device will comprise a time-of-propagation (TOP) counter in the
central part and a ring imaging Cherenkov system with a focusing aerogel
radiator (ARICH) in the forward region of the spectrometer. The TOP with
quartz radiator bars yields two-dimensional information from a Cherenkov
ring image based on the time of arrival and impact position of Cherenkov
photons. At a given momentum, the slower kaon (refer to the right plot
in Fig.~\ref{detect2}) emits Cherenkov photons at smaller angles than
the faster pion; as a result, the former photons also propagate slower
along the quartz bar. Both the barrel and endcap PID systems are expected
to considerably improve the hadron identification efficiency compared to
Belle. For instance, the ARICH alone will provide a $4\sigma$ $K$--$\pi$
separation up to the kinematic limits while the TOP counter can identify
kaons with an efficiency exceeding $90\%$ at a few percent pion
misidentification probability.

\section{Current Status and Schedule}

The SuperKEKB project received initial construction funding in 2010
for the positron damping ring, and a sizable fraction of funds (exceeding
100M\,US\$) under the Japanese `Very Advanced Research Support Program'
during the period 2010-2012 to upgrade the main rings. KEK hopes to
obtain additional funding to complete the construction as per schedule,
{\it i.e.}, start the SuperKEKB commissioning in early 2015 and begin
the data taking in late 2016. The commissioning itself will take place
in three phases: a) without final quads and Belle II detector, b) with
final quads and Belle II but no VXD, and c) with QCS and full detector.
It is expected that by 2019, the first $5\invab$ of data will be recorded
by Belle II while the full data sample of $50\invab$ will be reached in
2022-2023.

\section{Conclusions and Future Prospect}

Time and again, $e^+e^-$ $B$ factories have proven to be an excellent tool
for flavor physics producing a wealth of physics results; the most celebrated
one being the confirmation of the Kobayashi-Maskawa mechanism~\cite{km}
for $\CP$ violation in the SM. The ongoing upgrade of the KEKB accelerator
complex to SuperKEKB that plans to accumulate $50$ times more data will
take this legacy forward by providing a suitable probe for NP. Based on
a complementary approach to the energy-frontier experiments at LHC, Belle
II will focus on the study of rare decays of beauty and charm hadrons as
well as tau leptons. Particularly, a prolific charm physics program will be
a key component of the NP pursuit at Belle-II that includes an improved
sensitivity to charm mixing and $\CP$ violation, and a vigorous search for
rare or forbidden charm decays.


\Acknowledgements
We congratulate the organizers for an extremely stimulating workshop.
We are grateful to B. Golob, T. Hara, T. Kuhr, J. Libby, A. Schwartz,
K. Trabelsi, P. Urquijo, Y. Ushiroda and A. Zupanc for their helpful
comments and suggestions during the preparation of these proceedings.


\begin{thebibliography}{99}


\bibitem{Higgs}
S. Chatrchyan {\it et al.} (CMS Collaboration),
Phys.\ Lett.\ B {\bf 716}, 30 (2012);
G. Aad {\it et al.} (ATLAS Collaboration),
Phys.\ Lett.\ B {\bf 716}, 1 (2012).

\bibitem{cms}
S. Chatrchyan {\it et al.} (CMS Collaboration), JINST {\bf 3} (2008) S08004.

\bibitem{atlas}
G. Aad {\it et al.} (ATLAS Collaboration), JINST {\bf 3} (2008) S08003.

\bibitem{hou}
W.S. Hou, Int.\ J.\ Mod.\ Phys.\  D {\bf 20} (2011) 1521.

\bibitem{rubin}
V.C. Rubin and W.K. Ford, Jr., Astrophys.\ J.\ {\bf 159} (1970) 379.

\bibitem{lhcb}
A. Augusto Alves {\it et al.} (LHCb Collaboration), JINST {\bf 5}
(2010) P10003.

\bibitem{belle2}
T. Abe {\it et al.} (Belle II Collaboration), arXiv:1011.0352.

\bibitem{belle}
J. Brodzicka {\it et al.}, Prog.\ Theor.\ Exp.\ Phys., 04D001 (2012).

\bibitem{gim}
S. Glashow, J. Illiopoulos and L. Maiani, Phys.\ Rev.\ D {\bf 2},
1285 (1970).

\bibitem{scs}
Y. Grossman, A.L. Kagan and Y. Nir, Phys.\ Rev.\ D {\bf 75},
036008 (2007).

\bibitem{anze}
A. Zupanc {\it et al.} (Belle Collaboration), JHEP {\bf 09} (2013) 139.

\bibitem{origami}
C. Irmler {\it et al.}, Nucl.\ Instrum.\ Meth.\ A {\bf 732}, 109 (2013).

\bibitem{km}
M. Kobayashi and T. Maskawa, Prog.\ Theor.\ Phys. {\bf 49}, 652 (1973).

\end{thebibliography}
\end{document}